\documentstyle[11pt,newpasp,twoside]{article}
\def\edcomment#1{\iffalse\marginpar{\raggedright\sl#1\/}\else\relax\fi}
\reversemarginpar

\begin{document}
\title{Strange quark stars --- A review}
 \author{Renxin Xu}
\affil{School of Physics, Peking University, Beijing 100871,
       China}

\begin{abstract}

A pedagogical overview of strange quark matter and strange stars
is presented. After a historical notation of the research and an
introduction to quark matter, a major part is devoted to the
physics and astrophysics of strange stars, with attention being
paid to the possible ways by which neutron stars and strange stars
can be distinguished in astrophysics. Recent possible evidence for
{\em bare} strange stars is also discussed.

\end{abstract}

\section{Historical notes}

Soon after the Fermi-Dirac form (in 1926) of statistical mechanics
was proposed for particles which obey Pauli's exclusion principle
(in 1925), Fowler (1926) realized that the electron degeneracy
pressure can balance for those stars, called as white dwarfs,
discovered by astronomers in 1914.
By a numerical calculation (1931) for a polytropic gas of
extremely relativistic electrons, Chandrasekhar found a unique
mass, which was interpreted as a mass limit of white dwarfs.
Landau (1932) presented an elementary explanation of the
Chandrasekhar limit by considering the lowest total energy of
stars, and recognized that increasing density favors energetically
the formation of neutrons, discovered only several months before
by Chadwick, through the action
$p+e^-\leftrightarrow n+\nu_{\rm e}$.
A very massive object with much high density may have almost
neutrons in the chemical equilibrium, which was then called as
{\em neutron stars} (NSs).

Detailed calculations of NS structures showed (e.g., Oppenheimer
\& Volkoff 1939) that an NS can have a mass of $\sim 1 M_\odot$,
but is only $\sim 10$ km in radius, which makes it hard to be
observed by astronomers.
However, on one hand, a few authors do investigate possible
astrophysical implications of NSs. For example, Baade \& Zwicky
(1934) combined the researches of supernovae, cosmic rays, and
NSs, and suggested that NSs may form after supernovae; Pacini
(1967) even proposed that the stored energy in rotational form of
an NS could be pumped into the supernova remnant by emitting
electromagnetic waves.
On the other hand, NS models were developed with improved
treatments of equation of states, involving not only
\{$n,p,e^-$\}, but also mesons and hyperons. The cooling behavior
of NSs was also initiated in 1960s due to the discovery of X-ray
sources which were at first though mistakenly to be the thermal
emission from NSs.

The discovery of {\em radio pulsars} by Hewish \& Bell (and their
coauthors 1968) is a breakthrough in the study, and this kind of
stars were soon identified as spinning NSs by Gold (1968).
Since then more and more discoveries in other wave bands broaden
greatly our knowledge about these pulsar-like compact stars
(PLCSs), including X-ray pulsars, X-ray bursts, anomalous X-ray
pulsars, soft $\gamma$-ray repeaters, and ROSAT-discovered
``isolated neutron stars''.
It is still a current concept among astrophysicists that such
stars are really NSs.
NS studies are therefore in two major directions: 1, the emission
mechanisms for the stars, both rotation-powered and
accretion-powered; 2, the NS interior physics.

However, neutrons and protons are in fact {\em not} structureless
points although they were thought to be elementary particles in
1930s; they (and other hadrons) are composed of {\em quarks}
proposed by Gell-Mann and Zweig, respectively, in 1964.
The quark model for hadrons developed effectively in 1960s, and
Ivanenko \& Kurdgelaidze (1969) began to suggest a quark core in
massive compact stars. Itoh (1970) speculated about the exist of
3-flavor {\em full} quark stars (since only $u$, $d$ and $s$
quarks were recognized at that time), and even calculated the
hydrostatic equilibrium of such quark stars which are now called
as {\em strange stars} (SSs).
Is it possible that strange stars really exist in nature?
The possibility increases greatly if the Bodmer-Witten's
conjecture is correct: Bodmer (1971) initiated the discussion of
quark matter with lower energy per baryon than normal nucleus,
whereas Witten (1984) considered an assumption of stable 3-flavor
quark matter in details and discussion extensively three aspects
related (1, dark baryon and primordial gravitational wave due to
the cosmic separation in the QCD epoch; 2, strange quark stars; 3,
cosmic rays).
Farhi \& Jaffe's (1984) calculation in the MIT bag model showed
that the energy per baryon of strange matter is lower than that of
nucleus for QCD parameters within rather wide range although we
can hardly prove weather the Bodmer-Witten's conjecture is correct
or not from first principles.
Haensel, Zdunik \& Schaeffer (1986) and Alcock, Farhi \& Olinto
(1986) then modelled SSs, and found that SSs can also have typical
mass (of $\sim 1-2M_\odot$) and radius (of $\sim 10$ km), which
mean that {\em the pulsar-like compact stars believed previously
to be NSs might actually be SSs}.

Yet the most important and essential thing in the study is: how to
distinguish SSs from NSs observationally?
More and more SS candidates appeared recently in literatures
(e.g., Bombaci 2002, Xu 2002).
It is generally suggested that SSs as radio pulsars, the most
popular ones of PLCSs, should have crusts (with mass $\sim
10^{-5}M_\odot$) being similar to the outer crusts of NSs (Witten
1984, Alcock et al. 1986).
But this view was criticized by Xu \& Qiao (1998), who addressed
that {\em bare} strange stars (BSSs, i.e., SSs without crusts)
being chosen as the interior of radio pulsars have three
advantages: 1, the spectral features; 2, the bounding energy; and
3, the core collapse process during supernova. It is thus a new
window to distinguish BSSs from NSs via their magnetosphere and
surface radiation according to the striking differences between
the exotic quark surfaces of BSSs and the normal matter surfaces
of NSs.
With regard to the possible methods of finding strange stars in
literatures, hard evidence to identify a strange star may be found
by studying only the surface conditions since the other avenues
are subject to many complex nuclear and/or particle physics
processes that are poorly known.
Thanks to those advanced X-ray missions, it may be a very time for
us to identify real strange stars in the universe.

It is worth mentioning that, though some authors may name a
general term ``{\em neutron star}'', regardless of that the stars
are ``neutron'' or ``strange'', it is actually not suitable to
call a real SS to be an NS since no {\em neutron} in an SS.
%

\section{The standard model of particle physics and the quark matter}

One of the most great achievements in the last century is the
construction of the standard model in particle physics (e.g.,
Cottingham \& Greenwood 1998), which asserts that the material in
the universe is made up of elementary fermions (divided into
quarks and leptons) interacting though gauge bosons: photon
(electromagnetic), W$^\pm$ and Z$^0$ (weak), 8 types of gluons
(strong), and graviton (gravitational).
There are totally 62 types of ``building blocks'' in the model.
Besides the 13 types of gauge bosons, there are three generations
of fermions (1st: \{$\nu_{\rm e}$, e; u, d\}, 2nd: \{$\nu_{\rm
\mu}$, $\mu$; c, s\}, and 3rd: \{$\nu_{\rm \tau}$, $\tau$; t, b\}.
Note that each types of quarks has three colors) and their
antiparticles. The final one, which is still not discovered, is
the Higgs particle that is responsible to the origin of mass.

It is a first principle in the Yang-Mills theory that an
interaction satisfies a corresponding local gauge symmetry,
sometimes being broken spontaneously due to vacuum phase
transition.
The gauge theory for electromagnetic and weak interactions is very
successful, with a high precision in calculation; whereas one can
treat gravity using Einstein's general relativity theory if the
length scale is not as small as the Plank scale ($\sim 10^{-33}$
cm) although a gauge theory of gravity is still not successful.
As for the gauge theory for strong interaction, the quantum
chromodynamics (QCD), however, is still developing, into which
many particle physicists are trying to make efforts.

Nonetheless, QCD has two general properties. For strong
interaction in small scale ($\sim 0.1$ fm), i.e., in the high
energy limit, the interacting particles can be treated as being
{\em asymptotically free}; a perturbation theory of QCD (pQCD) is
possible in this case.
Whereas in larger scale ($\sim 1$ fm), i.e., in the low energy
limit, the interaction is very strong, which results in {\em color
confinement}. The pQCD is not applicable in this scale (many
non-perturbative effects appear then), and a strong interaction
system can be treated as a system of hadrons in which quarks and
gluons are confined. In this limit, we still have effective means
to study color interaction: 1, the lattice formulation (LQCD),
with the discretization of space-time and on the base of QCD,
provide a non-perturbative framework to compute numerically
relations between parameters in the standard model and experiments
by first principles; 2, phenomenological models, which rely on
experimental date available at low energy density, are advanced
for superdense hadronic and/or quark matter.

These two properties result in two distinct phases (Rho 2001) of
hadronic matter, depicted in the QCD phase diagram in terms of
temperature $T$ vs. baryon chemical potential $\mu_{\rm B}$ (or
baryon number density).
Hadron gas phase locates at the low energy-density limit where
both $T$ and $\mu_{\rm B}$ are relatively low, while a new phase
called {\em quark gluon plasma} (QGP) or {\em quark matter}
appears in the other limit when $T$ {\em or} $\mu_{\rm B}$ is high
although this new state of matter is still not found with certain
yet.
It is therefore expected that there is a kind of phase transition
from hadron gas to QGP (or reverse) at critical values of $T$ and
$\mu_{\rm B}$.
Actually a deconfinement transition is observed in numerical
simulations of LQCD for zero chemical potential $\mu_{\rm B}=0$,
when $T\rightarrow T_{\rm c}\simeq (150\sim 180)$ MeV.

Can we find real quark matter? Certainly we may improve very much
the knowledge about the strong interaction by studying the
matter's various properties if a QGP state is identified in hand.
One way is to create high energy-density fireball in laboratory
through the collisions of relativistic heavy ions in accelerators.
Quark matter is expected if the center of the fireball reaches a
temperature of $T_{\rm c}$, but the QGP is hadronized soon and
there are final states of hadrons detected.
It is thus a challenge to find clear signatures of QGP without
ambiguousness in these experiments.
Another way is to search celestial bodies which contain quark
matter via cosmological or stellar processes. Strange stars (\S 3)
are very probably such kind of objects, which are a kind of bulk
QGP with mass $\sim 1M_\odot$, composed of nearly equal number of
up, down, and strange quarks (called as strange quark matter, or
strange matter). Some possible observational signatures of SSs are
also addressed in literatures.
The first way to detect this new form of matter is in terrestrial
laboratory physics (lab-physics), whereas the second is in
astrophysics. They compensate each other.
Up to date, the research both in lab-physics and in astrophysics
faces a general difficulty in finding quark matter: to search a
{\em clear} signature for its existence.

In fact, the investigations in lab-physics and astrophysics are in
two {\em different} regions in the QCD phase diagram.
Available Monte Carlo simulations of LQCD are only applicable for
cases with $\mu_{\rm B}=0$, and thus give valuable guidance for
lab-physics.
However, it is a very different story with SSs, where the density
effects dominate ($\mu_{\rm B}\neq 0$).
It turns out, for technical reasons, to be extremely difficult to
study strange matter by LQCD, and we have to rely on
phenomenological models to speculate on the properties of SSs by
extrapolating our knowledge at nuclear matter density.
For an SS, with high $\mu_{\rm B}$ but low $T$, one can not
naively think that it is a simple QGP; in fact many interesting
phenomena, e.g., color superconductivity (Alford et al. 2001) of
strange matter, are discussed.
In this meaning, SSs provide only examples to study such hadronic
system at high density, and one thus has to learn ``experimental
date'' from astrophysics.

\section{Strange stars}

That strange stars (rather than neutron stars) are residual after
core-collapse type supernova explosion depends upon 1, quark
deconfinement occurs, 2, strange matter in bulk is absolutely
stable (Bodmer-Witten's conjecture). Unfortunately no certain
answer to these issues is obtained theoretically (\S 2).
Nevertheless, we still can obtain general concepts about those two
requirement for forming SSs by following simple arguments.
In the view of bag model\footnote{%
Quarks are though as degenerate Fermi gases in the model, which
exist only in a region of space endowed with a vacuum energy
density $B$ (called as the ``bag constant'').
}, %
the vacuum in and out hadrons is different; in-hadron is of ``pQCD
vacuum'' where the strong interaction is weak and pQCD is
applicable, but out-hadron is of ``QCD vacuum'' where the
nonlinear strong coupling between quarks or gluons is in control.
As baryon number density $\varrho$ gets higher and higher, the
pQCD vacuum in nucleons becomes more and more dominated; if
nucleon keeps a radius $r_{\rm n}\sim 1$ fm, the QCD vacuum
disappears when $\varrho>\varrho_{\rm c}\sim (4\pi r_{\rm
n}^3/3)^{-1}\sim 1.5\varrho_0$ ($\varrho_0$ is the density of
ordinary nuclear matter).
Depending on rotation frequency and stellar mass, such density are
easily surpassed in the cores of neutron stars (Weber 1999).
Therefore the first requirement -- quark deconfinement -- may be
satisfied.
In case of $\varrho>\varrho_{\rm c}$, $u$ and $d$ quarks
deconfined from nucleons could have a Fermi
energy\footnote{%
In the view of extremely relativistic ideal Fermi gases, one
nucleon (mainly neutron) contributes about 1 $u$ quark and 2 $d$
quarks, and the Fermi energy $\varepsilon_{\rm
F}=(3\pi^2)^{1/3}\hbar c\cdot n^{1/3}$ ($n$ the number density of
quarks). If $\varrho=1.5\varrho_0$, $\varepsilon_{\rm F}=379$ MeV
for $u$ quark and $\varepsilon_{\rm F}=477$ MeV for $d$ quark.
} %
$>\sim 400$ MeV, which is much larger than the mass of $s$ quark
$m_{\rm s}\sim 150$ MeV; the system should be energetically
favorable by opening of a third flavor degree of freedom ($s$
quark) since high-energy $u$ and $d$ quarks are expected to decay
into $s$ quarks via weak interactions.
The second requirement may thus also be reasonable.
It is worth addressing that, if only the first requirement is
satisfied while strange matter is not absolutely stable but only
metastable, a mixed phase where bulk quark and nuclear matter
could coexist over macroscopical distances may appear in NS cores
(such NSs are called as ``mixed stars'', see, e.g., Heiselberg \&
Hjorth-Jensen 2000, for a review).

These two requirements may lead to the existence of strange matter
in nature.
In a simplified version of the bag model, assuming quarks are
massless and noninteracting, we then have quark pressure $P_{\rm
q}=\rho_{\rm q}/3$ ($\rho_{\rm q}$ is the quark energy density);
the total energy density is $\rho=\rho_{\rm q}+B$ but the total
pressure is $P=P_{\rm q}-B$. One therefore have the equation of
state for strange matter,
\begin{equation}
P=(\rho-4B)/3.
\label{eos1}
\end{equation}
(Actually strange matter may have baryons from several
hundreds\footnote{%
Strange matter may have a minimum mass limit due to the surface
effects not being negligible, even if Bodmer-Witten's conjecture
is correct.
}, %
called as strangelets, to about that of our sun, called as strange
stars.)
One can obtain the mass $M$ and radius $R$ of an SS by integrating
numerically the TOV equation, with a strange matter equation of
state [e.g., eq.(1)], assuming a certain central density (or
pressure). The mass-radius relations of SSs calculated with
various strange matter equations of states are very different from
that of NSs, which may provide a possible way to differentiate SSs
from NSs (\S3.2). But an SS usually can have a mass of $1\sim
2M_\odot$, and correspondent radius $\sim 10$ km.

It is worth noting that the apparent temperature $T_\infty$ and
radius $R_\infty$ (also called as radiation radius, defined by
$L_\infty=4\pi R_\infty^2\sigma T_\infty^4$, where $L_\infty$ is
the apparent luminosity at infinity and $\sigma$ is the
Stefan-Boltzmann constant), i.e., the values observed at infinity,
is not that observed at the stellar surface ($R$ is also the
Schwartzschild radius coordinate at the surface, defined by
$\sqrt{{\rm surface\; area}/4\pi}$). For a compact star with mass
$M$ and radius $R$, the relations are (Haensel 2001)
$T_\infty = T \sqrt{1-(R_{\rm s}/R)}$, $R_\infty =
R/\sqrt{1-(R_{\rm s}/R)}$,
%
where $R_{\rm s}=2GM/c^2$ is the Schwartzschild radius.

\subsection{Structures: crusted or bare?}

As the strange quarks are more massive ($m_{\rm s}\sim 150$ MeV)
than the up ($m_{\rm u}\sim 5$ MeV) and down ($m_{\rm d}\sim 10$
MeV) quarks, some electrons are required to keep the chemical
equilibrium of an SS. This brings some interesting properties near
the bare quark surface.
Since quarks can be bound through strong interaction, whereas
electrons by much weaker interaction (only the electromagnetism),
the electrons can thus spread out the quark surface and be
distributed in such a way that a strong outward static electric
field is formed. Adopting a simple Thomas-Fermi model, one can
deduce analytical expressions for the electron number density
$n_{\rm e}$ and the electric field $E$ above the surface (Xu \&
Qiao 1999),
\begin{equation}
n_{\rm e} \sim {9.5 \times 10^{35} \over (1.2 z_{11} + 4)^3} \;\;
{\rm cm^{-3}},\;\; E \sim {7.2 \times 10^{18} \over (1.2 z_{11} +
4)^2} \;\; {\rm V\;cm^{-1},} %
\label{E}
\end{equation}
where $z$ is a measured height above the quark surface, $z_{11} =
z/(10^{-11}$ cm).
Very strong electric field, $\sim 10^{17}$ V/cm from eq.(2),
should be near the quark surface, which makes it possible to
support a normal-matter crust with mass $\sim 10^{-5}M_\odot$
(Alcock et al. 1986). SSs covered by such crusts are called as
crusted strange stars (CSSs), while SSs without crusts as bare
strange stars (BSSs).

Can radio pulsars, the largest population of PLCSs, be BSSs? The
answer was no in Alcock et al's (1986) paper: ``Pulsar emission
mechanisms which depend on the stellar surface as a source of
plasma will not work if there is a bare quark surface'' and ``The
universe is a dirty environment and a bare strange star may
readily accrete some ambient material''.
Their first point is certainly incomplete, because $e^\pm$ pairs
produced rapidly in strong electro-magnetic fields [as expressed
in eq.(2), or induced by the unipolar effect] should create a
corotating magnetosphere although no charged particles can be
pulled out into the magnetosphere (Xu \& Qiao 1998).
A nascent protostrange star should be bare because of strong mass
ejection and high temperature (Usov 1998) after the supernova
detonation flame; an SS can hardly accrete due to rapid rotating
and strong magnetic field (Xu et al. 2001); even accretion is
possible, a crust still can not form as long as the accretion rate
is not much larger than the Eddinton one (Xu 2002).
Their second point is thus also not reasonable.
Therefore radio pulsars could be BSSs. Furthermore, it is found
that BSSs may be better for explaining the observations of radio
pulsars as well as other PLCSs (\S3.3).

\subsection{Potential ways to identify a strange star}

Although the SS idea is not new, SS identification becomes a hot
topic only in recent years because of advanced techniques in
space.
Since SSs may have similar masses and radii, which are
conventional quantities observable in astronomy, to that of NSs,
it is very difficult to find observational signals of quark matter
in the PLCSs.
It was argued that the SS and NS cooling behaviors could be
distinguishable, since SSs may cool much faster than NSs (e.g.
Pizzochero 1991). However, recent more complete analyses on this
issue indicate that this may be impossible except for the first
$\sim$30 years after their births (Schaab et al. 1997).
Nevertheless, the author thinks there may still be three effective
ways to do.

1. The minimum rotation periods of SSs are smaller than that of
NSs.
Rotating stars composted of ideal fluid are subject to
rotation-mode instability, which leads to the loss of rotation
energy by gravitational radiation and results in substantial
spindown. However the matter of a real star is not ideal but have
viscosity; the calculated bulk viscosity, based on the work of
Wang \& Lu (1984), of strange matter is much higher than that of
neutron matter although their shear viscosities are similar;
therefore SSs could have smaller periods at which their higher
viscosity can prevent them from developing the instability (Madsen
1998, 2000).
The 2.14 ms optical source in SN 1987A (Middleditch  et al 2000)
should be an SS if being confirmed in further observations.

2. The approximate mass-radius ($M-R$) relations of SSs ($M\propto
R^3$) are in surprising contrast to that of NSs ($M\propto
R^{-3}$), and SSs can have much small radii.
Comparisons of observation-determined relations in X-ray binaries
with modelled ones may thus tell if an object is an NS or an SS
(Li et al. 1999). Also, a PLCS with radius $\la 8$ km could be an
SS (Drake et al. 2002).

3. There are striking differences between the surfaces of BSSs and
that of NSs. The very properties of the quark surface, e.g.,
strong bounding of particles, abrupt density change from $4B\sim
4\times 10^{14}$g/cm$^3$ to $\sim 0$ [eq.(1)] in $\sim 1$ fm, and
strong electric fields, may eventually help us to identify of a
BSS (\S3.3).

\subsection{Observational evidence for bare strange stars?}

Three parts of possible evidence for BSSs are discussed below.

Although pulsar emission mechanism is not well understood, the
RS-type (Ruderman \& Sutherland 1975) sparking model is still the
popular one to connect magnetospheric dynamics with general
observations, with an ``user friendly'' nature.
Maybe the strongest support to the RS-type vacuum gap model is the
drifting subpulses observed from some pulsars.
However RS model faces at least two difficulties for NSs: the
binding energy problem and the antipulsar issue, which can be
solved completely if radio pulsars are BSSs (Xu et al. 1999).

The soft $\gamma$-ray burst of SGR 0526-66, with peak luminosity
$\sim 10^7L_{\rm Edd}$, needs ultra-strong field ($\sim 10^{17}$G)
to constrain the fireball. An alternative bounding is through the
quark surface; and it may be natural to explain the bursting
energy and the light curves in a framework that a comet-like
object falls to a BSS (Zhang et al. 2000, Usov 2001).

BSSs are expected to have featureless spectra (of both surface
thermal and magnetospheric non-thermal components) since no ion is
above the quark surface or in the magnetosphere, except for
electron cyclotron lines due to the Landau levels appeared in
strong fields (Xu \& Qiao 1998, Xu 2002).
Recent observations known hitherto of several PLCSs actually show
featureless spectra except for two sources 1E1207 and SGR1806, the
lines of which may originate from Landau level transitions in
suitable field strength for the space facilities
(astro-ph/0207079).

\section{Conclusions}

The theoretical bases of SSs are, to some extent, solid in physics
and the formation of strange matter stars is possible in
astrophysics; SSs could thus exist.
Although each of the observed phenomena from PLCSs may be
interpreted under the NS regime with unusual or artificial
physical properties, it could be a natural way to understand the
observations by updating NSs with SSs.


\noindent
{\em Acknowledgements}. I wish to thank Profs. Ignazio Bombaci,
Chongshou Gao, Naoki Itoh, Jes Madsen, and Vladimir Usov for their
comments and discussions.


\begin{references}

\reference{Alcock, C., Farhi, E., \& Olinto, A. 1986, ApJ, 310,
261}

\reference{Alford, M., Bowers, J. A., Rajagopal, K. 2001, in
Lecture Notes in Physics, 578, Physics of Neutron Star Interiors,
ed. D. Blaschke, et al. (Springer), 235}

\reference{Baade, W., \& Zwicky, F. 1934, Phys. Rev., 45, 138}

\reference{Bodmer, A. R. 1971, Phys. Rev., D4, 1601}

\reference{Bombaci, I. 2002, invited talk, Compacta stars in the
QCD phase diagrams, Copenhagen (Denmark), eConf C010815, 29
(astro-ph/0201369)}

\reference{Chandrasekhar, S. 1931, ApJ, 74, 81}

\reference{Cottingham, W. N., \& Greenwood, D. A. 1998, {\em An
introduction to the standard model of particle physics}, Cambridge
Univ. Press}

\reference{Drake, J. J., et al. 2002, ApJ, 572, 996}

\reference{Farhi, E., \& Jaffe, R. L. 1984, Phys. Rev., D30, 2379}

\reference{Fowler, R. H. 1926, MNRAS, 87, 114}

\reference{Gold, T. 1968, Nature, 218, 731}

\reference{Haensel, P. 2001, A\&A, 380, 186}

\reference{Haensel, P., Zdunik, J. L., \& Schaeffer, R. 1986,
A\&A, 160, 121}

\reference{Heiselberg, H., Hjorth-Jensen, M. 2000, Phys. Rep.,
328, 237}

\reference{Hewish, A., Bell, S.J., Pilkington, et al. 1968,
Nature, 217, 709}

\reference{Itoh N. 1970, Prog. Theor. Phys., 44, 291}

\reference{Ivanenko, D., \& Kurdgelaidze, D. F. 1969, Lett. Nuovo
Cimento, 2, 13}

\reference{Landau, L. D., 1932, Phys. Z. Sowjetunion, 1, 285 (an
expanded discussion of his ideal can be found in one of their
courses ``{\em Statistical Physics}''.)}

\reference{Li, X. D., et al. 1999, Phys. Rev. Lett., 83, 3776}

\reference{Madsen, J. 1998, Phys. Rev. Lett., 81, 16}

\reference{Madsen, J. 2000, Phys. Rev. Lett., 85, 10}

\reference{Middleditch, J. et al. 2000, New Astronomy, 5, 243}

\reference{Oppenheimer, J. R., \& Volkoff, G. M. 1939, Phys. Rev.,
55, 374}

\reference{Pacini, F. 1967, Nature, 216, 567}

\reference{Pizzochero, P. M. 1991, Phys. Rev. Lett., 66, 2425}

\reference{Rho, M. 2001, in AIP Conf. Proc., 556, Explosive
phenomena in astrophysical compact objects, ed. H.-Y. Chang et al.
(New York: AIP), 160}

\reference{Ruderman, M.A., \& Sutherland, P.G. 1975, ApJ, 196, 51}

\reference{Schaab, C., Hermann, B., Weber, F., \& Weigel, M.K.
1997a, ApJ, 480, L111}

\reference{Usov, V. V. 1998, Phys. Rev. Lett., 80, 230}

\reference{Usov, V. V. 2001, Phys. Rev. Lett. 87, 021101}

\reference{Wang, Q. D., \& Lu, T. 1984, Phys. Lett., B148, 211}

\reference{Weber, F. 1999, J. Phys. G: Nucl. Part. Phys., 25,
R195}

\reference{Witten, E., 1984, Phys. Rev., D30, 272}

\reference{Xu, R.X. 2002, ApJ, 570, L65}

\reference{Xu, R. X., \& Qiao, G. J. 1998, Chin. Phys. Lett., 15,
934}

\reference{Xu, R. X., \& Qiao, G. J. 1999, Chin. Phys. Lett., 16,
778}

\reference{Xu, R.X., Qiao, G.J., Zhang, B., 1999, ApJ, 522, L109}

\reference{Xu, R. X., Zhang, B., Qiao, G. J. 2001, Astropart.
Phys., 15, 101}

\reference{Zhang, B., Xu, R. X, Qiao, G. J. 2000, ApJ, 545, L127}

\end{references}
\end{document}